\documentclass[journal,1pt,doublecolumn]{IEEEtran}
\usepackage{mathpple}
\usepackage{times}
\usepackage{dsfont}
\usepackage{amsmath}  
\usepackage{amssymb}  
\usepackage{mathrsfs} 
\usepackage{theorem}  
\usepackage{cite}     
\usepackage{comment}  
\usepackage[colorlinks,
			linkcolor=black,
			anchorcolor=black,
			citecolor=black
			]{hyperref}
\usepackage{upref}
\usepackage{amsfonts}
\usepackage{dsfont}
\usepackage{verbatim}
\usepackage{algorithm}
\usepackage{algorithmic}
\usepackage{bm}
\usepackage[dvipsnames,usenames]{color}
\usepackage{enumerate}

\usepackage{graphicx}
\usepackage{subfigure}

\usepackage{latexsym}

\usepackage{color}
\usepackage{multirow}
\usepackage{footnote}
\usepackage{booktabs}
\usepackage{cite}
\usepackage{hhline}
\usepackage{gensymb}
\usepackage{tikz}
\usetikzlibrary{shapes.geometric}
\usetikzlibrary{arrows.meta,arrows}
\usepackage{graphicx}
\usepackage{xcolor}
\usepackage{listings}
\usepackage[
top    = 0.75in,
bottom = 1.05in,
left   = 0.63in,
right  = 0.63in]{geometry}
\usepackage{caption}
\usepackage{romannum}

\newcommand{\Code}[2]{y_{\{#1,#2\}}}

\newcommand{\hatp}[1]{\hat{P}_{#1}}
\newcommand{\Hx}{H(\mathbf{X})}
\newcommand{\tmin}{\text{min}}

\newcommand{\be}[1]{\begin{equation}\label{#1}}
\newcommand{\ee}{\end{equation}}

\newcommand{\qed}{\hfill$\Box$\\[1ex]}

\newcommand{\Cref}[1]{Co\-rol\-la\-ry\,\ref{#1}}

\theoremstyle{plain} \theorembodyfont{\normalfont\slshape}

\newtheorem{thm}{Theorem$\!$}
\newenvironment{theorem}{\begin{thm}\hspace*{-1ex}{\bf.}}{\end{thm}}

\newtheorem{prop}[thm]{Proposition$\!$}

\newtheorem{lem}[thm]{Lemma$\!$}
\newenvironment{lemma}{\begin{lem}\hspace*{-1ex}{\bf.}}{\end{lem}}

\newtheorem{cor}[thm]{Corollary$\!$}

\newtheorem{prob}[thm]{Problem$\!$}

\newtheorem{defi}[thm]{Definition$\!$}

\theorembodyfont{\normalfont}

\newtheorem{exam}{Example$\!$}
\newenvironment{example}{\begin{exam}\hspace*{-1ex}{\bf .}}{\end{exam}}

\newtheorem{remrk}{Remark$\!$}
\newenvironment{remark}{\begin{remrk}\hspace*{-1ex}{\bf .}}{\end{remrk}}

\definecolor{Codecolor}{named}{White}  

\begin{document}
\title{Rate-Constrained Shaping Codes for Finite-State Channels with Cost\vspace{-.2em}}


\author{
	\IEEEauthorblockN{
		\textbf{Yi~Liu}~\IEEEmembership{Member,~IEEE,}\IEEEauthorrefmark{1}
		\textbf{Yonglong~Li}~\IEEEmembership{Member,~IEEE,}\IEEEauthorrefmark{1}\IEEEauthorrefmark{3} 
		\textbf{Pengfei~Huang}~\IEEEmembership{Member,~IEEE,}\IEEEauthorrefmark{1}\\ 
		\textbf{Paul~H.~Siegel}~\IEEEmembership{Life Fellow,~IEEE}\IEEEauthorrefmark{1} 	 
	}
	
	\IEEEauthorblockA{
		\IEEEauthorrefmark{1}Department of Electrical and Computer Engineering, CMRR, University of California, San Diego, La Jolla, CA
	}
	\IEEEauthorblockA{
		\IEEEauthorrefmark{3}National University of Singapore
	}
%
%
	
yiliu0812@gmail.com, elelong@nus.edu.sg, pehuangucsd@gmail.com, psiegel@ucsd.edu\vspace{-5ex}
}

\maketitle
\pagenumbering{gobble}
\thispagestyle{plain}
\pagestyle{plain}
\begin{abstract}
	%
	
	Shaping codes are used to generate code sequences in which the symbols obey a prescribed probability distribution.  They arise naturally in the context of source coding for noiseless channels with unequal symbol costs.  Recently, shaping codes have been proposed to extend the lifetime of flash memory and reduce DNA synthesis time. In this paper, we study a general class of shaping codes for noiseless finite-state channels with cost and i.i.d. sources.  We establish a relationship between the code rate and minimum average symbol cost.  We then determine the rate that minimizes the average cost per source symbol (total cost).  An equivalence is established between codes minimizing average symbol cost and codes minimizing total cost, and a separation theorem is proved, showing that optimal shaping can be achieved by a concatenation of optimal compression and optimal shaping for a uniform i.i.d. source.  
	\vspace{-3.5ex}
\end{abstract}

\section{Introduction}
\label{sec::intro}
Shaping codes are used to encode information for use on channels with symbol costs under an average cost constraint. They find application in data transmission with a power constraint, where constellation shaping is achieved by addressing into a suitably designed multidimensional constellation or, equivalently, by incorporating, either explicitly or implicitly, some form of non-equiprobable signaling. More recently, shaping codes have been proposed for use in data storage applications:  coding for flash memory to reduce device wear~\cite{LiuSieGC16}, and  coding for efficient DNA synthesis in DNA-based storage~\cite{Lenz2020}. Motivated by these applications,~\cite{ShapingJournal} investigated  information-theoretic properties and design of rate-constrained fixed-to-variable length shaping odes for memoryless noiseless  channels with cost and general i.i.d. sources. In this paper, we extend the results in~\cite{ShapingJournal} to rate-constrained shaping codes for finite-state  noiseless  channels with cost and general i.i.d. sources.

Finite-state noiseless  channels with cost trace their conceptual origins to Shannon's 1948 paper that launched the study of information theory~\cite{Shannon}. In that paper, Shannon considered the problem of transmitting information over a telegraph channel. The telegraph channel is a finite-state graph and the channel symbols -- dots and dashes -- have different time durations, which can be interpreted as integer transmission costs. Shannon defined the combinatorial capacity of this channel and gave an explicit formula. He also determined the symbol probabilities that maximize the entropy per unit cost, and showed the equivalence of this probabilistic definition of capacity to the combinatorial capacity. In~\cite{Csiszar1969}, this result was then generalized to arbitrary non-negative symbol costs. In \cite{Khandekar}, a new proof technique  for deriving the combinatorial capacity was introduced for non-integer costs and another proof of the equivalence of combinatorial and probabilistic definitions of capacity was given. In~\cite{BochererThesis} and~\cite{BochererSCC}, a generating function approach was used to extend the equivalence to a larger class of constrained systems.   

We refer to the problem of designing codes that achieve the capacity, i.e., that maximize the information rate per unit cost, or, equivalently, that minimize  the cost per information bit, as the  \textit{type-\Romannum{2} coding problem}. 
Several researchers have considered this problem.  In~\cite{Csiszar1969},  modified Shannon-Fano codes,  based on matching the probability of source and codeword sequences, were introduced, and they were shown to be asymptotically optimal. A similar  idea was used in~\cite{SavariGallager}, where an arithmetic coding technique was introduced. 
Several works extend  coding algorithms for memoryless channels to finite-state channels. In~\cite{BochererThesis}, a finite-state graph was transformed to its memoryless representation $\mathcal{M}$ and a normalized geometric Huffman code was used to design a asymptotically capacity achieving code on $\mathcal{M}$. In~\cite{Iwata}, the author extended the dynamic programming algorithm introduced in~\cite{Golin} to finite-state channels. The proposed algorithm finds locally optimal codes for each starting state, but the algorithm does not guarantee global optimality. In~\cite{Fujita}, an iterative algorithm that can find globally optimal codes was proposed.

The concepts of combinatorial capacity and probabilistic capacity can be generalized to the setting where there is a constraint on the average cost per transmitted channel symbol.  The probabilistic capacity was determined in ~\cite{McElieceRodemich} and~\cite{Justesen}, where the entropy-maximizing stationary Markov chain satisfying the average cost constraint was found.  
The relationship between  cost-constrained combinatorial capacity and probabilistic capacity was also addressed in~\cite{Karabed}.
The equivalence of the two definitions of cost-constrained capacity was proved in~\cite{SoriagaITA}, and an alternative proof was recently given in~\cite{Lenz2021}, where methods of analytic combinatorics in several variables were used to directly evaluate the cost-constrained combinatorial capacity.   



We refer to the problem of designing codes that achieve the cost-constrained capacity as the  \textit{type-\Romannum{1} coding problem}.  This problem  has also been addressed by several authors. In~\cite{Karabed}, an asymptotically optimal block code was introduced by considering codewords that start and end at the same state.  
In~\cite{Khayrallah}, the authors construct fixed-to-fixed length and variable-to-fixed length codes based on state-splitting methods~\cite{ACH} for  magnetic recording and constellation shaping applications.  
Other constructions can be found in~\cite{Krachkovsky},~\cite{Soriaga} and~\cite{SoriagaISIT}. 

In this paper, we  address the problem of designing  shaping codes for noiseless finite-state channels with cost and general i.i.d. sources.    We systematically study the fundamental properties of these codes from the perspective of symbol distribution, average cost, and entropy rate using the theory of finite-state word-valued sources. We derive fundamental bounds relating these quantities and  establish an equivalence between optimal \textit{type}-\Romannum{1} and  \textit{type}-\Romannum{2} shaping codes.  
A generalization of Varn coding~\cite{Varn} is shown to provide an asymptotically optimal \textit{type}-\Romannum{2} shaping code for uniform i.i.d. sources. Finally, we prove separation theorems showing that optimal shaping for a general  i.i.d. source can be achieved by a concatenation of optimal lossless compression with an optimal shaping code for a uniform i.i.d. source.  

In Section~\ref{state::sec::finitestatechannel}, we define finite-state channels with cost and review the combinatorial and probabilistic capacities associated with the \textit{type}-\Romannum{1} and  \textit{type}-\Romannum{2} coding problems. 
In Section~\ref{state::sec::preliminary}, we define finite-state variable length shaping codes for  channels with cost and characterize properties of the codeword process using the theory of finite-state word-valued sources. 
In Section~\ref{state::sec::main}, we analyze shaping codes for a fixed code rate, which we call \textit{type}-\Romannum{1} \textit{shaping codes}. We develop a theoretical bound on the trade-off between the rate -- or more precisely, the corresponding \emph{expansion factor} -- and the average cost  of a type-\Romannum{1} shaping code. We then study shaping codes that minimize average cost per source symbol (total cost). We refer to this class of shaping codes as \textit{type-\Romannum{2} shaping codes}. We derive the relationship between the code expansion factor and the total cost and determine the optimal expansion factor.
In Section~\ref{state::sec::design}, we consider the problem of designing optimal shaping codes. We prove an equivalence theorem showing that both type-\Romannum{1} and type-\Romannum{2} shaping codes can be realized using a type-\Romannum{2} shaping code for a channel with modified edge cost. Using a generalization of Varn coding~\cite{Varn}, we propose an asymptotically optimal type-\Romannum{2} shaping code on this modified channel for a uniform i.i.d. source. We then extend our construction to arbitrary i.i.d. sources by introducing a separation theorem, which states that optimal shaping can be achieved by a concatenation of lossless compression and optimal shaping for a uniform i.i.d. source.

Due to space constraints, we must omit many detailed proofs, which can be found in~\cite{YiThesis}. However, we remark that several new proof techniques are required to  extend  the results on  block shaping codes for  memoryless channels in~\cite{ShapingJournal}   to the  corresponding results on finite-state shaping codes for finite-state channels  in this paper. 

\vspace{-1.5ex}
\section{Noiseless Finite-State Costly Channel}

\label{state::sec::finitestatechannel}
Let $\mathcal{H} = (\mathcal{V},\mathcal{E})$ be an irreducible  finite  directed graph, with vertices $\mathcal{V}$  and edges $\mathcal{E}$. A \textit{finite-state costly channel} is a noiseless channel with cost associated with $\mathcal{H}$, where each edge $e\in \mathcal{E}$ 
is assigned a non-negative cost $w(e)\geq 0$. We assume that between any  pair of vertices $(v_i,v_j)\in \mathcal{V}\times \mathcal{V}$, there is at most one edge. If not, we can always convert it to another graph that satisfies this condition by state splitting~\cite{Marcus}. An example of such a channel is given in Example~\ref{state::example::flash_memory_channel}.

\begin{figure}
	\centering
	\resizebox{!}{0.4\columnwidth}{
	\begin{tikzpicture}[->,>=stealth',shorten >=1pt,auto,node distance=1.5cm,thick,main node/.style={circle,draw,font=\Large\bfseries,minimum size=0.5cm}]
	
	\node[main node] (1) {00};
	\node[main node] (2) [below left of=1] {01};
	\node[main node] (3) [below right of=1] {10};
	\node[main node] (4) [below right of=2] {11};
	
	\path (1) edge [loop above] node {1} (1) edge node [left] {2} (2)
	(2) edge node [left] {4} (4) edge [bend right] node {4} (3)
	(3) edge node [right] {2} (1) edge [bend right] node {3} (2)
	(4) edge [loop below] node {4} (4) edge node [right] {4} (3);
	\end{tikzpicture}
}
	\caption{Flash memory channel}
	\label{state::fig::flash_memory_channel}
	\end{figure}
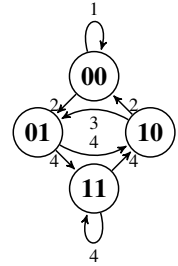

\vspace{-1.5ex}
\begin{example}
	\label{state::example::flash_memory_channel}
	In   SLC NAND flash memory, cells are arranged in a grid and programming a cell affects its neighbors. One example of this phenomenon is \textit{inter-cell inference} (ICI)~\cite{Taranalli}. Cells have two states: \textit{programmed}, corresponding to bit 1, and \textit{erased}, corresponding to bit 0. Due to ICI, programming a cell will damage its neighbors cells. Each length-3 sequence has a cost associated with the damage to the middle bit, as shown in Table~\ref{state::table::slcflashchannelcost}. 	We can convert this table into a directed graph with vertices $\mathcal{V} = \{00,01,10,11\}$, as shown in Fig.~\ref{state::fig::flash_memory_channel}.
	
	 \vspace{0.5ex}
	\begin{table}[h]
	\caption{Flash memory channel cost}
		\centering
		\begin{tabular}{@{}ccccccccc@{}}
			\toprule
			$e$   & 000 & 001 & 010 & 011 & 100 & 101 & 110 & 111 \\ \midrule
			$w(e)$&   1 &   2 &   4 &   4 &   2 &   3 &   4 &   4 \\\bottomrule
		\end{tabular}
	\label{state::table::slcflashchannelcost}
	 \vspace{-0.5ex}
	\end{table}
\end{example}

	\vspace{-4.0ex}
\subsection{Channel capacity with average cost constraint}
\label{state::subsec::typeonechannel}
Given a length-$n$ edge sequence $e_1^n$, the cost of this sequence is defined as $
W(e_1^n) = \sum_{i=1}^n w(e_i),$
and the average cost of this sequence is defined as $
A(e_1^n) = \frac{1}{n} W(e_1^n)$.
If  $K_n(W)$ is the number of sequences of length-$n$ with average cost less than or equal to $W$, then the \textit{combinatorial capacity}  for a given average cost constraint~\cite{SoriagaITA}, or \textit{cost-constrtained capacity},  is
\vspace{-1ex}
\begin{equation}
C_{\Romannum{1}, \text{comb}}(W) \mathrel{\overset{\makebox[0pt]{\mbox{\normalfont\tiny\sffamily def}}}{=}} \limsup_{n\rightarrow \infty} \frac{1}{n} \log_2 |K_n(W)|.
\vspace{-0.5ex}
\end{equation}
We also refer to this definition as \textit{type-\Romannum{1} combinatorial capacity}. 

Let $\mathbf{E}$ be a stationary Markov process with entropy rate $H(\mathbf{E})$ and average cost $A(\mathbf{E})$. The \textit{probabilistic capacity} for a given average cost constraint $W$, or \textit{cost-constrained probabilistic capacity},  is  
\vspace{-1ex}
\begin{equation}
C_{\Romannum{1}, \text{prob}}(W) \mathrel{\overset{\makebox[0pt]{\mbox{\normalfont\tiny\sffamily def}}}{=}} \sup_{\mathbf{E}: A(\mathbf{E})\leq W}H(\mathbf{E}).
\vspace{-0.5ex}
\end{equation}
The maxentropic Markov chain for a given $W$ was derived in~\cite{Justesen} and~\cite{McElieceRodemich}. The result relies on the one-step cost-enumerator matrix $ D(S)$, where $S\geq 0$, with entries
\vspace{-1ex}
\begin{equation}
\label{state::equ::introd}
d_{ij}(S) = \begin{cases}
2^{-Sw(e_{ij})} \quad \text{for edge $e_{ij}$ between $(v_i,v_j)$}\\
0 \quad \text{if edge $e_{ij}$ doesn't exist.}
\end{cases}
\vspace{-0.5ex}
\end{equation}
Denote by $\lambda(S)$ its Perron root and by vectors $E_L = [P_{v_i} / \rho_i]$ and $E_R = [\rho_i]^\top$ the corresponding left and right eigenvectors such that $E_LE_R = 1$. Given an average cost constraint $W(S)$ 
the maxentropic Markov chain has transition probabilities
\begin{equation}
 P_{ij}(S) = \frac{1}{\rho_i} \frac{2^{-S w(e_{ij})}}{\lambda(S)} \rho_j 
 \end{equation}
such that
\vspace{-1ex}
\begin{equation}
W(S) = \frac{1}{\lambda(S)}\sum_{ij}P_{v_i}2^{-Sw(e_{ij})}\frac{\rho_j}{\rho_i},
\vspace{-0.5ex}
\end{equation}
and the type-\Romannum{1} probabilistic capacity of this channel is
\vspace{-1ex}
\begin{equation}
C_{\Romannum{1}, \text{prob}}(W(S)) = \log_2 \lambda(S) + SW(S).
\vspace{-0.3ex}
\end{equation}
It was shown in~\cite{SoriagaITA},\cite{Lenz2021} that $C_{\Romannum{1}, \text{comb}}(W) = C_{\Romannum{1}, \text{prob}}(W)$.
\vspace{-2.5ex}
\subsection{Channel capacity without cost constraint}
\label{state::subsec::typetwochannel}
Denote by $K(W)$ the number of distinct sequences $e^*$ with cost  equal to $W$. The \textit{combinatorial capacity}, or the \textit{type-\Romannum{2} combinatorial capacity}, of this channel is defined as
\vspace{-1ex}
\begin{equation}
C_{\Romannum{2}, \text{comb}}\mathrel{\overset{\makebox[0pt]{\mbox{\normalfont\tiny\sffamily def}}}{=}} \limsup_{W\rightarrow \infty} \frac{1}{W}\log_2 K(W).
\vspace{-0.5ex}
\end{equation}
Similarly, the \textit{type-\Romannum{2} probabilistic capacity} of this channel is defined as
\vspace{-1ex}
\begin{equation}
C_{\Romannum{2}, \text{prob}} \mathrel{\overset{\makebox[0pt]{\mbox{\normalfont\tiny\sffamily def}}}{=}} \sup_{\mathbf{E}}\frac{H(\mathbf{E})}{A(\mathbf{E})}.
\vspace{-0.5ex}
\end{equation}
In~\cite{Khandekar}, it was proved that the transition probabilities of the maxentropic Markov process are $P_{ij}(S_0)$, 
where $S_0$ satisfies $\lambda(S_0) = 1$. It was also proved that
\vspace{-1ex}
\begin{equation}
C_{\Romannum{2}, \text{comb}} = C_{\Romannum{2}, \text{prob}} = S_0.
\vspace{-0.5ex}
\end{equation}
In~\cite{BochererThesis} and~\cite{BochererSCC}, the equivalence between $C_{\Romannum{2}, \text{comb}}$ and $C_{\Romannum{2}, \text{prob}}$ was extended to a larger class  of constrained systems.

\section{Finite-State Variable-Length Codes: A Word-Valued Source Approach}
\label{state::sec::preliminary}
\subsection{Finite-State Variable-Length Codes}
\begin{figure}
		\centering
	\resizebox{!}{0.2\columnwidth}{
	\begin{tikzpicture}[->,>=stealth',shorten >=0.5pt,auto,node distance=1.5cm,thick,main node/.style={circle,draw,font=\Large\bfseries}]
	
	\node[main node] (1) {$v_0$};
	\node[main node] (2) [right of=1] {$y_1$};
	\node[main node] (3) [right of=2] {$v_1$};
	\node[main node] (4) [right of=3] {$y_2$};
	\node[main node] (5) [right of=4] {$v_2$};
	\node[main node] (6) [above of=2] {$x_1$};
	\node[main node] (7) [above of=4] {$x_2$};
	
	\path (1) edge  (2)
	(2) edge (3)
	(3) edge (4)
	(4) edge (5)
	(6) edge (2)
	(7) edge (4);
	
	\end{tikzpicture}
}
	\caption{Encoding Process.}
	\label{state::fig::encoding_process}
\end{figure}
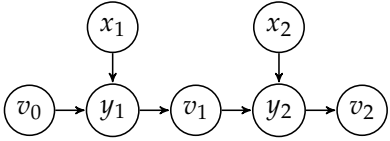

\vspace{-0.5ex}
Let $\mathbf{X} = X_1X_2\ldots$, where $X_i  \sim X$ for all $i$,  be an i.i.d. source with alphabet $\mathcal{X}=\{\alpha_1,\ldots,\alpha_u\}$. We denote by $P_i$ the probability of symbol $\alpha_i$ and assume that $P_1 \geq P_2 \geq \ldots \geq P_u$.  Let$|\mathcal{X}|$   denote the size of the alphabet and  $P(x^*)$  denote the probability of any finite sequence $x^*$. A finite-state variable-length code on graph $\mathcal{H}$ is a mapping {$\phi: V\times\mathcal{X}^q  \rightarrow \mathcal{E}^*$.} For simplicity and without loss of generality,  we assume $q = 1$ and   denote $\phi(v_i, x_j)$ as $\Code{i}{j}$. The starting state of $\Code{i}{j}$ is $v_i$ and is denoted by $start(\Code{i}{j})$. Its ending state is denoted by $end(\Code{i}{j})$. Its length is denoted by $L(\Code{i}{j})$. We assume this mapping has the following two properties:

\begin{itemize}
	\item Its subcodebook set $\mathcal{Y}_i = \{\Code{i}{j}\}$ is prefix-free for all $v_i$.
	\item $l \leq L(\Code{i}{j}) < L$ for some positive $l$ and $L$.
\end{itemize}
The encoding process associated with this mapping, as shown in Fig.~\ref{state::fig::encoding_process}, is defined as follows.
\begin{itemize}
	\item Start: fixed state $v_0$.
	\item Input: source word sequence $x_1,x_2,x_3,\ldots$; $x_i \in \mathcal{X}^q$.
	\item Output: codeword sequence $y_1,y_2,y_3,\ldots$;  $y_i \in \mathcal{E}^*$.
	\item Encoding rules: $y_i = \phi(v_{i-1},x_i),\, v_{i} = end(y_i).$
\end{itemize}

From the encoding rules, we have
\vspace{-1ex}
\begin{equation}
\begin{aligned}
y_i &= \phi(y_{i-1},x_i) = \phi(end(y_{i-1}), x_i) = F(y_{i-1}, x_i),\\
v_{i}& = end(y_i) = end(\phi(v_{i-1},x_i)) = G(v_{i-1}, x_i).
\end{aligned}
\vspace{-0.5ex}
\end{equation}
These equations suggest that we can define \textit{codeword graph} $\mathcal{F}_0$ and \textit{state graph} $\mathcal{G}_0$, which are closely related to the encoding process, as follows.
\begin{itemize}
	\item Codeword graph $\mathcal{F}_0 = (\mathcal{V}_{F_0}, \mathcal{E}_{F_0})$:
	\begin{itemize}
		\item Vertices $\mathcal{V}_{F_0}=\{\Code{i}{j}\text{ for all $i$ and $j$}\}$.
		\item Edges $\mathcal{E}_{F_0}=\{e = (\Code{i}{j},\Code{k}{l}) | v_k =  end(\Code{i}{j})\}$.
	\end{itemize}
	\item State graph $\mathcal{G}_0 = (\mathcal{V}_{G_0}, \mathcal{E}_{G_0})$:
	\begin{itemize}
		\item Vertices $\mathcal{V}_{G_0}=\mathcal{V}$.
		\item Edges $\mathcal{E}_{F_0}=\{e = (v_i ,v_j) | \exists x_k \text{ s.t. } v_j = G(v_i,x_k))\}$.
	\end{itemize}
\end{itemize}

Here we construct an irreducible subgraph from $\mathcal{F}_0$ and $\mathcal{G}_0$. We choose the irreducible component $\mathcal{G} \subseteq\mathcal{G}_0$ that contains $v_0$ (we assume that $v_0$ always belongs to one of the irreducible components). Its state and edge sets are denoted by $\mathcal{V}_G$ and $\mathcal{E}_G$ respectively. For convenience, we will re-index the vertices in $\mathcal{V}_G$ as $v_0,v_1,v_2,\ldots$. The encoding process is a Markov process associated with $\mathcal{G}$ and transition probabilities
\vspace{-1ex}
\begin{equation}
t^G_{ij} = \sum_{k:G(v_i,x_k)=s_j} P(x_k).
\vspace{-0.5ex}
\end{equation}
We denote by $\pi_G(v_i)$ the stationary distribution of $v_i$.
We define a subgraph $\mathcal{F}\subset \mathcal{F}_0$ as follows. Its vertex $\mathcal{V}_F$ is
\vspace{-1ex}
\begin{equation}
\mathcal{V}_F = \{\Code{i}{j} = \phi(v_i, x_j) |\text{ for } v_i \in \mathcal{V}_G\text{ and all $x_j$} \}
\vspace{-0.5ex}
\end{equation}
and its edge set $\mathcal{E}_F$ is
\vspace{-1ex}
\begin{equation}
\begin{aligned}
\mathcal{E}_F  = \{ &(\Code{i}{j},y_{\{k,l\}}) | \Code{i}{j},y_{\{k,l\}} \in \mathcal{V}_F \\ &\text{ and }v_k = end(\Code{i}{j}) = end( \phi(v_i, x_j))\}.
\end{aligned}
\vspace{-0.5ex}
\end{equation}
We can prove the following lemma.
\vspace{-1.5ex}
\begin{lemma}
	\label{state::lemma::graph_equivalence}
	The graph $\mathcal{F}$ is irreducible. The encoding process is a Markov process associated with $\mathcal{F}$ with transition probabilities
	\vspace{-0.5ex}
	\begin{equation}
	P(y_{\{k,l\}} | \Code{i}{j}) = P(x_l).
	\vspace{-0.5ex}
	\end{equation}
	The stationary distribution of states $\{\Code{i}{j}\}$ is
	\vspace{-0.5ex}
	\begin{equation}
	\pi_{F}(\Code{i}{j}) =\pi_G(v_i) P(x_j).
	\vspace{-0.5ex}
	\end{equation}\qed
\end{lemma}
\vspace{-2.5em}
Using the law of large numbers for irreducible Markov chain~\cite[Exercise~5.5]{Durrett},~\cite[Theorem 3.21]{Marcus} and the dominated convergence theorem, we know that the expected length of the codeword process $\mathbf{E}$ is 
\vspace{-0.5ex}
\begin{equation}
\label{state::equ::el}
\small{\begin{aligned}
\mathbb{E}(L)  &\mathrel{\overset{\makebox[0pt]{\mbox{\normalfont\tiny\sffamily def}}}{=}} \lim_{n\rightarrow \infty}\frac{1}{n} \mathbb{E}(\sum_{m=1}^{n} L(Y_m)) =\sum_{ij} L(\Code{i}{j}) \pi_G(v_i) P(x_j).
\end{aligned}}
\vspace{-0.5ex}
\end{equation}

Given an edge $e\in \mathcal{E}$ and codeword $\Code{i}{j}$, we denote by $N_{e}(\Code{i}{j})$ the total number of occurrences of $e$ in sequence $\Code{i}{j}$. We  can similarly prove that 
\begin{equation}
\begin{aligned}
\mathbb{E}(N_{e}) &= \sum_{\Code{i}{j}} N_{e}(\Code{i}{j}) \pi_G(v_i) P(x_j).
\end{aligned}
\end{equation}

\vspace{-2.5ex}
\subsection{Finite-State Word-Valued Source}
Introduced in~\cite{NishiaraMorita}, a \textit{word-valued source} is a discrete random process that is formed by sequentially encoding the symbols of an i.i.d. random process $\mathbf{X}$ into corresponding codewords over an alphabet $\mathcal{E}$. 
In this paper, the mapping is a function of both input symbols and the starting state. We refer to the process $\mathbf{E}$, formed by an i.i.d. source process and mapping $\phi$, as a \emph{finite-state word-valued source}. 

Given an encoded sequence $e_1e_2\ldots$, the probability of sequence $e_1^n$ is $Q(e_1^n)$, and the number of occurrences of $e$ is $N_e(e_1^n)$. The following properties of the process $\mathbf{E}$ are of interest.
\begin{itemize}
	\item The asymptotic symbol occurrence probability
	\vspace{-0.5ex}
	\begin{equation}
	\hatp{e}= \lim_{n\rightarrow\infty} \frac{1}{n} \mathbb{E}(N_{e}(E_1^n)).
	\vspace{-0.5ex}
	\end{equation}
	\item The asymptotic average cost
	\vspace{-0.5ex}
	\begin{equation}
	A(\phi) = \lim_{n\rightarrow \infty} \frac{1}{n} \mathbb{E}(W(E_1^n)).
	\vspace{-0.5ex}
	\end{equation}
	\item The entropy rate
	\vspace{-0.5ex}
	\begin{equation}
	H(\mathbf{E}) = \lim_{n\rightarrow \infty} \frac{1}{n} H(E_1^n).
	\end{equation}
\end{itemize}

We can prove the following lemma.
 \vspace{-1ex}
\begin{lemma}
	\label{state::lemma:asymptotic_occurrence}
	For a finite-state code {$\phi:\mathcal{V}\times \mathcal{X}^q \rightarrow \mathcal{E}^*$} associated with graph $\mathcal{F}$ and stationary distribution $\{\pi_F{\Code{i}{j}} = \pi_G(v_i) P(x_j)\}$ such that $\mathbb{E}(N_{e}) < \infty$ for all $e\in \mathcal{E}$ and ${\mathbb{E}(L)<\infty}$, the asymptotic probability of occurrence $\hatp{e}$ is given by 
	 \vspace{-0.5ex}
	\begin{equation}
	\hatp{e}=\mathbb{E}(N_{e})\frac{1}{\mathbb{E}(L)}.
	\vspace{-0.5ex}
	\end{equation} 
	The asymptotic average cost $A(\phi)$ is 
	\vspace{-0.5ex}
	\begin{equation}
	A(\phi) = \sum_{e\in \mathcal{E}} \hatp{e}w(e).
	\vspace{-0.5ex}
	\end{equation}
	\vspace{-3em}~\qed
\end{lemma}
\vspace{-1.5ex}
\begin{lemma}
	\label{state::lemma::entropyrate}
	For a finite-state  code $\phi:\mathcal{V}\times \mathcal{X}^q \rightarrow \mathcal{E}^*$ associated with graph $\mathcal{F}$ such that $H(\mathbf{X}) < \infty$ and $\mathbb{E}(L) < \infty$, the entropy rate of the codeword process is 
	\vspace{-1ex}
	\begin{equation}
	H(\mathbf{E})  = \frac{q\Hx}{\mathbb{E}(L)} = \frac{\Hx}{f}.
	\vspace{-0.5ex}
	\end{equation} 
	Here $f =\mathbb{E}(L)/q$ is the \textit{expansion factor} of the mapping $\phi$.~\qed
	\vspace{-2.5em}
\end{lemma}

\subsection{Asymptotic normalized KL-divergence}
	\vspace{-0.5ex}
Similar to the definition of $\hatp{e}$, the asymptotic probability of occurrence of state $v\in\mathcal{V}$ is defined as
	\vspace{-1ex}
\begin{equation}
\hatp{v} = \lim_{n\rightarrow\infty} \frac{1}{n} \mathbb{E}(N_{v}(E_1^n)),
	\vspace{-0.8ex}
\end{equation}
and we can prove that
	\vspace{-1ex}
\begin{equation}
\label{state::equ::distribution_equation}
\hatp{v_i} = \sum_{j} \hatp{e_{ij}}
	\vspace{-0.8ex}
\end{equation}
Consider a finite-order Markov process $\hat{\mathbf{E}}$ associated with graph $\mathcal{H}$ and transition probabilities
	\vspace{-1ex}
\begin{equation}
t^H_{ij} = \frac{\hatp{e_{ij}}}{\hatp{v_i}}.
	\vspace{-0.8ex}
\end{equation}
Denote by $\hat{P}(e_1^n)$ the probability of a length-$n$ sequence generated by this process. To measure the difference between $\hat{\mathbf{E}}$ and $\mathbf{E}$, we define the asymptotic normalized KL-divergence as
\vspace{-1ex}
\begin{equation}
\lim_{n\rightarrow \infty} \frac{1}{n}D(E_1^n || \hat{E}_1^n) = \lim_{n\rightarrow \infty}\sum_{e_1^n\in \mathcal{E}^n} Q(e_1^n) \log_2 \frac{Q(e_1^n)}{\hat{P}(e_1^n)}. 
	\vspace{-0.8ex}
\end{equation}
The relationship between processes $\mathbf{E}$ and $\hat{\mathbf{E}}$ is summarized in the following lemma.
	\vspace{-2.2ex}
\begin{lemma}
	\label{state::lemma::entropy_rate_bound}
	The asymptotic normalized KL-divergence between processes $\mathbf{E}$ and $\hat{\mathbf{E}}$ satisfies
	\begin{equation}
	\lim_{n\rightarrow \infty} \frac{1}{n}D(E_1^n || \hat{E}_1^n) = H(\hat{\mathbf{E}}) - H(\mathbf{E}) = H(\hat{\mathbf{E}}) - \frac{\Hx}{f}.
	\vspace{-0.5ex}
	\end{equation}\vspace{-1ex}\qed
	\vspace{-2 em}
\end{lemma}

\begin{remark}
	\label{state::rmk::iid}
	When $H(\hat{\mathbf{E}}) = H(\mathbf{E})$, $\lim_{l\rightarrow \infty}\frac{1}{l}D(E_1^n|| \hat{E}_1^n)=0$. Therefore, the codeword process $\mathbf{E}$ approximates the stationary Markov process $\hat{\mathbf{E}}$, in the sense that the asymptotic normalized KL-divergence between $\mathbf{E}$ and $\hat{\mathbf{E}}$ converges to 0.
		\vspace{-1.8ex}
\end{remark}

\section{Optimal Shaping Codes for Finite-State Costly Channel}
\label{state::sec::main}
	\vspace{-0.5ex}
In this section, we first analyze shaping codes that minimize the average cost with a given expansion factor. 
We refer to this minimization problem as the \textit{type-\Romannum{1} shaping problem}, and we call shaping codes that achieve the minimum average cost for a given expansion factor \textit{optimal type-\Romannum{1} shaping codes}. We solve the following optimization problem.
\vspace{-0.5ex}
\begingroup
\small
\begin{equation}
\begin{aligned}
  \underset{\hatp{e_{ij}}}{\text{minimize}\;\;\;}
  &\sum_{ij}\hatp{e_{ij}} w(e_{ij})\\
 \text{subject to \;\;\;}
& H(\hat{\mathbf{E}}) = -\sum_{ij} \hatp{e_{ij}} \log_2 \frac{\hatp{e_{ij}}}{\sum_j \hatp{e_{ij}}}\geq \frac{\Hx}{f}\\
&  \sum_{j} \hatp{e_{ji}} = \sum_{j}\hatp{e_{ij}}   {\;  \; \rm and  \;\; }
  \sum_{ij}\hatp{e_{ij}}=1.
\end{aligned}
\vspace{-0.5ex}
\end{equation}
\endgroup
In~\cite{Lenz2021}, the authors discuss  \textit{cost-diverse} and \textit{cost-uniform} graphs. A graph is cost-diverse if it has at least one pair of equal-length paths with different costs that connect the same pair of vertices.  Otherwise it is called cost-uniform. It can be proved that the edge costs  $w(e_{ij})$ of a cost-uniform graph can be expressed as $w(e_{ij}) = -\mu_i + \mu_j - \alpha$. The following theorem relates to the achievable minimum average cost of a finite-state shaping code.
\vspace{-1.5ex}
\begin{theorem}
	\label{state::thm::type1_shaping}
	On a cost-diverse graph, the average cost of a {type-\Romannum{1}} shaping code $\phi: \mathcal{V}\times\mathcal{X}^q  \rightarrow \mathcal{E}^*$ with expansion factor $f$ is lower bounded by 
	\vspace{-1ex}
	\begingroup
	\small
	\begin{equation}
	A_\tmin(f) = \sum_{ij}\hatp{e_{ij}} w(e_{ij}) = \frac{\Hx}{Sf} -\frac{\log_2 \lambda(S)}{S},
	\vspace{-0.5ex}
	\end{equation}
	\endgroup
	\vspace{0.05ex}
	where $\small
	\hatp{e_{ij}}=\frac{\hatp{v_i}}{\rho_i} \frac{2^{-Sw(e_{ij})}}{\lambda(S)} \rho_j,$
	$\lambda(S)$ is the Perron root of the matrix  $D(S)$,
	$E_L = [\hatp{v_i}/ \rho_i]$ $E_R = [\rho_i]^\top$ are the corresponding eigenvectors such that $E_LE_R = 1$, and $S$ is the constant such that
	\vspace{-1ex}
	\begin{equation}
	H(\hat{\mathbf{E}}) = -\sum_{ij} \hatp{e_{ij}} \log_2 \frac{\hatp{e_{ij}}}{\hatp{v_i}} = H(\mathbf{E})= \frac{\Hx}{f}.
	\vspace{-0.5ex}
	\end{equation}
	On a cost-uniform graph, the average cost for any shaping code is a constant $-\alpha$.\qed
	\vspace{-5.5ex}
\end{theorem}
\begin{remark}
	\label{state::rmk::approxiamatemarkov}
	When the minimum average cost is achieved, we have $H(\hat{\mathbf{E}})= H(\mathbf{E})$. As shown in Remark~\ref{state::rmk::iid}, the codeword sequence approximates a finite-order stationary Markov process with transition probabilities $\{\hatp{e_{ij}}/ \hatp{v_i}\}$.\qed
	\vspace{-4.5ex}
\end{remark}

Using Theorem~\ref{state::thm::type1_shaping}, we study shaping codes that minimize average cost per source symbol, which are closely related to the type-\Romannum{2} channel capacity introduced in Section~\ref{state::subsec::typetwochannel}. The \textit{total cost} of a shaping code is
\vspace{-1ex}
\begingroup
\small
\begin{equation}
\begin{aligned}
T(\phi)& = \lim_{n\rightarrow \infty}\frac{\mathbb{E}(W(\phi(X^{nq})))}{nq}  = f\sum_{ij} \hatp{e_{ij}}w(e_{ij}).
\end{aligned}
\vspace{-0.5ex}
\end{equation}
\endgroup
We refer to the problem of  minimizing the total cost as the \textit{type-\Romannum{2} shaping problem}. Shaping codes that achieve the minimum total cost are referred to as \textit{optimal type-\Romannum{2} shaping codes}. The corresponding optimization problem is as follows.
\vspace{-1ex}
\begingroup
\small
\begin{equation}
\begin{aligned}
  \underset{\hatp{e_{ij}},f}{\text{minimize}\;\;\;}  
&  f\sum_{ij} \hatp{e_{ij}}w(e_{ij})\\
  \text{subject to \;\;\;} 
 & H(\hat{\mathbf{E}})\geq H(\mathbf{E}) = \frac{\Hx}{f}\\
&    \sum_j \hatp{e_{ij}} = \sum_j \hatp{e_{ji}} {\;  \; \rm and  \;\; }
  \sum_{ij}\hatp{e_{ij}}=1.
\end{aligned}
\vspace{-1ex}
\end{equation} 	
\endgroup
We have the following theorem that determines the minimum achievable total cost of a shaping code.
\vspace{-1ex}
\begin{theorem}
	\label{state::thm::type2_shaping}
	If  a cost-0 cycle does not exist,  the minimum total cost of a type-\Romannum{2} shaping code $\phi: \mathcal{V}\times\mathcal{X}^q  \rightarrow \mathcal{E}^*$ is given by 
	\vspace{-1ex}
	\begin{equation}
	T_\tmin = f^{\star}\sum_{ij}\hatp{e_{ij}}^{\star} w(e_{ij}) = \frac{\Hx}{S^{\star}},
		\vspace{-0.5ex}
	\end{equation}
	where $\hatp{e_{ij}}^{\star}=\frac{\hatp{v_i}^{\star}}{\rho_i} 2^{-S^{\star} w(e_{ij})}\rho_j,$ $S^{\star}$ is a constant such that  $\lambda(S^{\star}) = 1$, and
	$E_L = [\hatp{v_i}^{\star} / \rho_i]$ and $E_R = [\rho_i]^\top$ are the corresponding eigenvectors such that $E_LE_R = 1$. The corresponding expansion factor $f^{\star}$ is
	\vspace{-1ex}
	\begingroup
	\small
	\begin{equation}
	f^{\star}= \frac{\Hx}{-\sum_{ij} \hatp{e_{ij}}^{\star} \log_2 \frac{\hatp{e_{ij}}^{\star} }{\hatp{v_i}^{\star}}} = \frac{\Hx}{S^{\star}\sum_{ij}\hatp{e_{ij}}^{\star} w(e_{ij})}.
	\vspace{-0.5ex}
	\end{equation}
	\endgroup
	If there is a cost-0 cycle in $\mathcal{H}$, the total cost is a decreasing function of $f$.\qed
	\vspace{-6.5ex}
\end{theorem}

\section{Optimal Shaping Code Design}
\vspace{-0.5ex}
\label{state::sec::design}
In this section, we consider the problem of designing optimal type-\Romannum{1} and type-\Romannum{2} shaping codes.
\vspace{-2.5ex}
\subsection{Equivalence Theorem}
We consider the channel with modified edge costs
\vspace{-1ex}
\begin{equation}
w'(e_{ij}) = -\log_2 \frac{\hatp{e_{ij}}^{\star}}{\hatp{v_i}^{\star}},
\vspace{-0.5ex}
\end{equation}
where $\hatp{e_{ij}}^{\star}, \hatp{v_i}^{\star}$ are given in Theorem~\ref{state::thm::type2_shaping}. 
It is easy to check that the optimal type-\Romannum{2} shaping codes on this channel are also optimal on the original channel, in the sense that the symbol occurrence probabilities \{$\hatp{e_{ij}}$\} are identical on both channels. We can prove the following lemma.
\vspace{-2ex}

\begin{lemma}
	\label{state::lemma::equivalence_type2}
	Given a noiseless finite-state costly channel with edge costs $\{w(e_{ij})\}$. If there is a shaping code $\phi: \mathcal{V}\times\mathcal{X}^q  \rightarrow \mathcal{E}^*$ such that
	\vspace{-1ex}
	\begin{equation}
	f\sum_{ij}\hatp{e_{ij}}w'(e_{ij}) - \Hx < \delta,
	\vspace{-0.5ex}
	\end{equation}
	where 	$w'(e_{ij}) = -\log_2 (\hatp{e_{ij}}^{\star} / \hatp{v_i}^{\star}) = S^{\star} w(e_{ij}) + \log_2 \rho_i - \log_2 \rho_j,$
  for some $\delta > 0$, then the total cost of this code satisfies
	\vspace{-1ex}
	\begin{equation}
	f\sum_{ij}\hatp{e_{ij}}w(e_{ij}) - \frac{\Hx}{S^{\star}} < \frac{\delta}{S^{\star}}.
	\vspace{-0.5ex}
	\end{equation}
	\vspace{-1ex}\qed
 \vspace{-4.5ex}
\end{lemma}
The next two theorems establish the equivalence between type-\Romannum{1} and type-\Romannum{2} shaping codes. 
\vspace{-1ex}
\begin{theorem}
	\label{state::thm::equivalence_type1}
	Given a noiseless finite-state costly channel with edge costs $\{w(e_{ij})\}$. For any $\gamma, \eta > 0$, there exists a $\delta >0$ such that if there exists a shaping code $\phi: \mathcal{V}\times\mathcal{X}^q  \rightarrow \mathcal{E}^*$ with expansion factor $f'$ such that
	\vspace{-0.5ex}
	\begin{equation}
	f'\sum_{ij}\hatp{e_{ij}}'w'(e_{ij}) - \Hx < \delta,\vspace{-0.5ex}
	\end{equation} 
	where 	$w'(e_{ij}) {=} -\log_2 \frac{\hatp{e_{ij}}^{\star}}{\hatp{v_i}^{\star}} = S^{\star} w(e_{ij}) + \log_2 \rho_i - \log_2 \rho_j,$
	then the average cost of this code is upper bounded by
	\vspace{-1ex}
	\begin{equation}
	\sum_{ij}\hatp{e_{ij}}'w(e_{ij}) - (\frac{\Hx}{Sf} - \frac{\log_2\lambda(S)}{S} ) < \gamma
	\vspace{-0.5ex}
	\end{equation}
	and the expansion factor of this code $f'$ satisfies $|f' - f |< \eta$.
	\vspace{-2ex}
\end{theorem}

\begin{theorem}
	\label{state::thm::equivalence_type1to2}
	Given a noiseless finite-state costly channel that does not contain a cost-0 cycle. Denote by $S^{\star}$ the constant such that $\lambda(S^\star) = 1$ and by $f^{\star}$ the expansion factor of an optimal type-\Romannum{2} shaping code. For any $\gamma > 0$, there exist  $\delta ,\eta>0$ such that if a shaping code $\phi: \mathcal{V}\times\mathcal{X}^q  \rightarrow \mathcal{E}^*$ with expansion factor $f'$ satisfies
	\vspace{-1ex}
	\begin{equation}
	\sum_{ij}\hatp{e_{ij}}'w(e_{ij}) - A_\tmin(f') < \delta,\quad |f'-f^{\star}|<\eta,
	\vspace{-0.5ex}
	\end{equation}
	then the total cost of this code satisfies
	\vspace{-1ex}
	\begin{equation}
	f'\sum_{ij}\hatp{e_{ij}}'w(e_{ij}) - \frac{\Hx}{S^{\star}} < \gamma.
	\vspace{-0.5ex}
	\end{equation}
	\vspace{-5ex}
\end{theorem}

\subsection{Generalized Varn Code}
\vspace{-0.5ex}
\label{state::subsec::varn}
We now describe  an asymptotically optimal type-\Romannum{2} shaping code for uniform i.i.d. sources based on a generalization of Varn coding~\cite{Varn}.
Given a uniform i.i.d. input source $\mathcal{X}$, a \textit{generalized Varn Code} on the noiseless finite-state costly channel is a collection of tree-based variable-length mappings, $\phi: \mathcal{V}\times\mathcal{X}^q  \rightarrow \mathcal{E}^*$. Denote by $\mathcal{Y}_k$ the set of codewords starting from state $v_k$, namely
\vspace{-1ex}
\begin{equation}
\mathcal{Y}_k = \{\phi(v_k, x^q)| x^q \in \mathcal{X}^q\}.
\vspace{-0.5ex}
\end{equation}
Codewords in $\mathcal{Y}_k$ are generated according to the following steps.
\begin{itemize}
	\item Set state $v_k\in\mathcal{V}$ as the root of the tree.
	\item Expand the root node. The edge costs \{$w'(e_{kl})\}$ are the modified costs defined in Lemma~\ref{state::lemma::equivalence_type2}. The cost of a leaf node is the cost of the path from root node to the leaf node.
	\item Expand the leaf node that has the lowest cost.
	\item Repeat the previous steps until the total number of leaf nodes $M \geq |\mathcal{X}|^q$. Delete the leaf nodes that have the largest cost until the number of leaf nodes equals to $|\mathcal{X}|^q$. Each path from the root node $v_k$ to a leaf node represents one codeword in $\mathcal{Y}_k$.
\end{itemize}
The following lemma gives an upper bound on the total cost of a generalized Varn code.
\vspace{-1.5ex}
\begin{lemma}
	\label{state::lemma::generalized_varn}
	The total cost of a generalized Varn code $\phi: \mathcal{V}\times\mathcal{X}^q  \rightarrow \mathcal{E}^*$ is upper bounded by
	\begingroup
	\small
	\begin{equation}
	\begin{aligned}
	T(\phi) &\leq \frac{\log_2 M}{q} + \frac{\max_{ij}\{w'(e_{ij})\}}{q} \underset{q \rightarrow \infty}{=} \log_2 |\mathcal{X}|.
	\end{aligned}
	\end{equation} 
	\endgroup\qed
		\vspace{-5ex}
\end{lemma}
\begin{remark}
	By extending some leaf nodes to states that are not visited by the original code, we can make graph $\mathcal{G}_0$ a complete graph. Then we can choose any state as the starting state. This operation only adds a constant to the cost of a codeword and therefore does not affect the asymptotic performance of the generalized Varn code.
\end{remark}
\vspace{-3ex}
\begin{example}
	For the channel introduced in Example~\ref{state::example::flash_memory_channel}, the optimal symbol distributions that minimize the total cost are shown in Table~\ref{state::table::SLCoptimaldistribution}. Based on the distribution,
	we can design a generalized Varn code  on the channel with modified edge costs shown in Table~\ref{state::tab::flashmodifiedcost}. 
	The total cost as a function of codebook size is shown in Fig.~\ref{state::fig::generalizedvarn_originalchannel}.

\vspace{1em}		
		\begin{table}[h]
		\caption{Probabilities for SLC flash channel that minimize total cost.}    
		\centering
		\resizebox{0.95\columnwidth}{!}{
			\begin{tabular}{@{}ccccccccc@{}}
				\toprule
				$u$       & 000   & 001   & 010   & 011   & 100   & 101   & 110   & 111   \\ \midrule
				$\hatp{u}$& 0.4318& 0.1323& 0.1135& 0.0593& 0.1323& 0.0405& 0.0593& 0.0310\\\bottomrule
			\end{tabular}
		}
		\label{state::table::SLCoptimaldistribution}
	\end{table}
\begin{table}[h]
\caption{Modified cost for flash memory channel.}    
	\centering
	\resizebox{0.95\columnwidth}{!}{
		\begin{tabular}{@{}ccccccccc@{}}
			\toprule
			$u$   & 000  & 001  & 010  & 011  & 100  & 101  & 110  & 111 \\ \midrule
			$C(u)$&0.3805&2.0923&0.6068&1.5423&0.3855&2.0923&0.6068&1.5423 \\\bottomrule
		\end{tabular}
	}
	\label{state::tab::flashmodifiedcost}
\end{table}
	\begin{figure}[h]
		\centering
		\includegraphics[width=0.95\columnwidth]{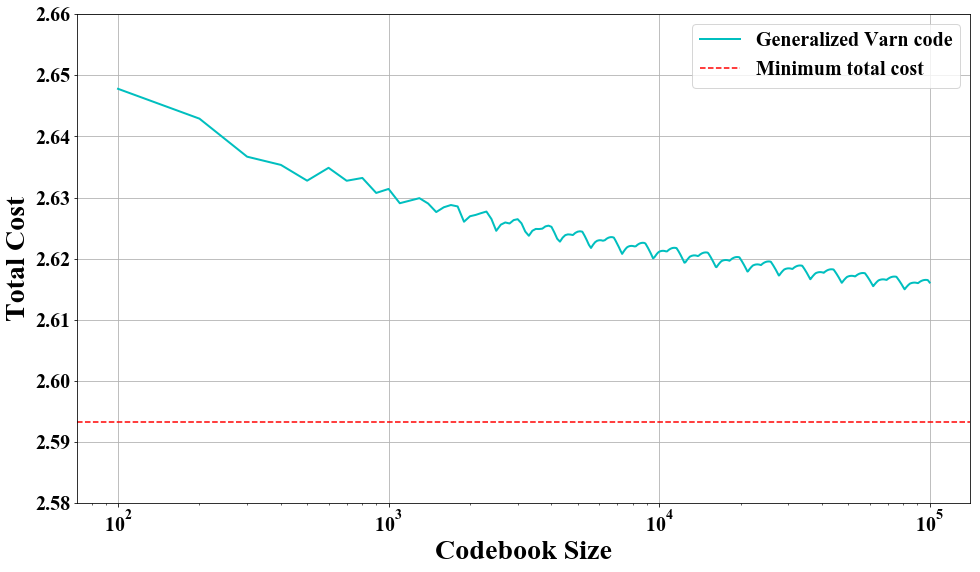}
		\caption{The total cost of a generalized Varn code on the SLC flash channel}
		\label{state::fig::generalizedvarn_originalchannel}
			\vspace{-2.5ex}
	\end{figure}
\end{example}

\subsection{Separation Theorem}
\vspace{-0.5ex}
We now present a separation theorem for shaping codes. It states that the minimum total cost can be achieved by a concatenation of optimal lossless compression with an optimal shaping code for a uniform i.i.d. source.
\begin{theorem}
	\label{state::thm::separation_type2}
	Given an i.i.d. source $\mathbf{X}$ and a noiseless finite-state costly channel with edge costs $\{w(e_{ij})\}$, the minimum total cost can be achieved by a concatenation of an  optimal lossless compression code with a binary optimal type-\Romannum{2} shaping code for a uniform i.i.d. source.
\end{theorem}

\begin{theorem}
	\label{state::thm::separation_type1}
	Given the i.i.d. source $\mathbf{X}$, the noiseless finite-state costly channel with edge costs $\{w(e_{ij})\}$, and the expansion factor $f$, the minimum average cost can be achieved by a concatenation of an optimal lossless compression code with a binary optimal type-\Romannum{1} shaping code for uniform i.i.d. source and expansion factor $f' = \frac{f}{\Hx}.$
\end{theorem}
By Theorem~\ref{state::thm::equivalence_type1to2}, the optimal type-\Romannum{1} shaping code for uniform i.i.d. source in Theorem~\ref{state::thm::separation_type1} can be replaced by a suitable  optimal type-\Romannum{2} shaping code for uniform i.i.d. source.

\end{document}